%% file: main.tex
\documentclass[conference]{IEEEtran}
\input{settings}

\usepackage{setspace} 
\def\smallerspacecaption{\vspace{-1mm}}

\begin{document}

\title{TrojanSAINT: Gate-Level Netlist Sampling-Based Inductive Learning for Hardware Trojan Detection}
\input{authors}

\maketitle
\renewcommand{\headrulewidth}{0.0pt}
\thispagestyle{fancy}
\lhead{}
\rhead{}
\chead{\copyright~2023 IEEE.
This is the author's version of the work.
The definitive Version of Record is published in the IEEE International Symposium on Circuits and Systems (ISCAS), 2023. }
\cfoot{}

\begin{abstract}
\input{abstract}
\end{abstract}


\section{Introduction}
\label{sec:intro}

Integrated circuit (IC) design and manufacturing has become an increasingly outsourced process that
involves various third parties. While this has allowed to increase both productivity and complexity of ICs, it has
also made them more vulnerable to the introduction of hardware Trojans (HTs), among other threats.
HTs are malicious circuitry, causing
system failure, leaking sensitive information, etc~\cite{rajendran2013high, karri2010trustworthy}. Thus, methods to accurately check for HTs become increasingly important.

Conventional methods for HT detection include code review~\cite{yasaei2021gnn4tj} and verification against a ``golden
reference'', i.e., a trusted, HT-free version of the design~\cite{faezi2021htnet}.
However, the former is prone to errors, especially for complex ICs, and the latter is not always feasible,
especially when untrusted parties are engaged in the design process~\cite{faezi2021htnet}. Other methods
have been proposed as well, e.g., utilizing side-channel
fingerprinting~\cite{he2018golden}; however, such
are limited to post-silicon HT detection. Researchers have shown that machine learning (ML) can successfully adapt to a wide variety of HTs, without necessitating new techniques for detecting new HT designs~\cite{hasegawa2017trojan}.

Using graph neural networks
(GNNs) is an emerging and promising method toward this end~\cite{yasaei2021gnn4tj,yu2021hw2vec,GNN_security_survey,alrahis2022graph}.
Thanks to their ability to work on graph-structured data -- such as circuits --
GNNs can leverage both a) the features of each gate and b) the overall structure of the
design for the prediction of HTs.

Still, prior art for GNN-based HT detection suffers from the following limitations (also summarized in Table~\ref{tab:comparison}).

\begin{table}[tb]
\centering
\caption{Comparison of GNN-based HT Detection Schemes}
\label{tab:comparison}
\resizebox{\columnwidth}{!}{%
\renewcommand\arraystretch{1.2}
\begin{tabular}{cccccc}
\hline
\multirow{2}{*}{\textbf{Method}} & \textbf{HT} & \textbf{HT} & \textbf{Gate-Level} & \textbf{Pre-} & \textbf{Post-}\\ 
& \textbf{Detection} & \textbf{Localization} & \textbf{Netlist} & \textbf{Silicon} & \textbf{Silicon}\\ \hline
GNN4TJ~\cite{yasaei2021gnn4tj} & Yes & No & No & Yes & No \\\hline
HW2VEC~\cite{yu2021hw2vec} & Yes & No & No & Yes & No\\\hline
GRFTL~\cite{yasaei2022golden} & Yes & Yes & No & Yes & No \\\hline

\textbf{Our Work} & \textbf{Yes} & \textbf{Yes} & \textbf{Yes}& \textbf{Yes} & \textbf{Yes}\\\hline
\end{tabular}%
}
\end{table}

\begin{figure}[tb]
\centerline{\includegraphics[width=.95\columnwidth]{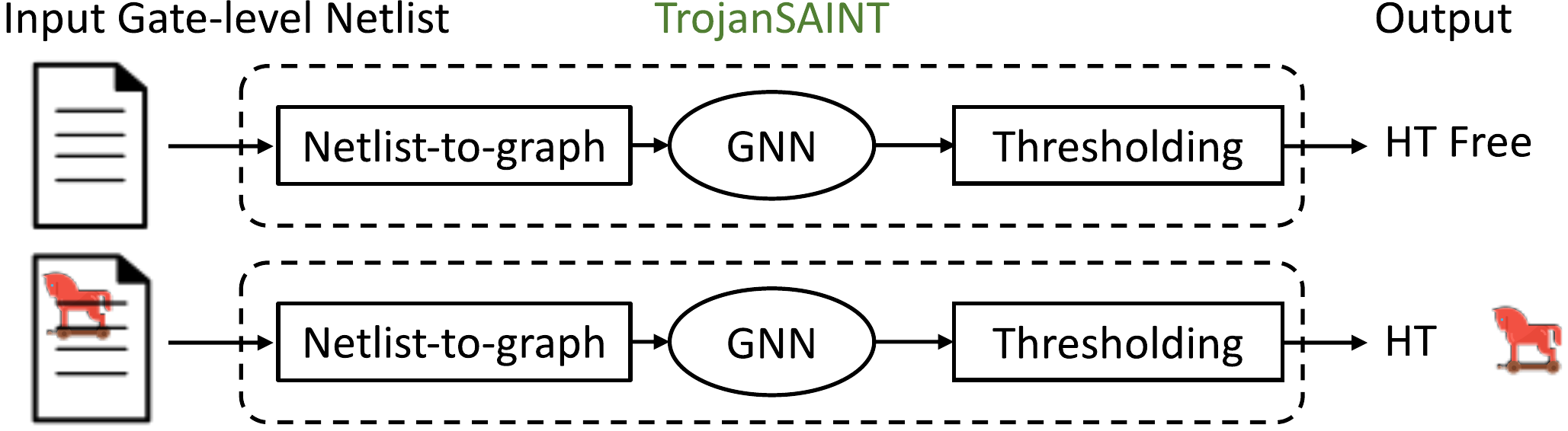}}
\caption{Concept of TrojanSAINT.}
\label{fig:intro}
\end{figure}

\textbf{HT Localization.}
State-of-the-art GNN-based detection schemes, GNN4TJ~\cite{yasaei2021gnn4tj} and
HW2VEC~\cite{yu2021hw2vec}, predict whether a design contains a HT or not, but they cannot localize HTs. However,
localizing HTs is essential to identify the part of the design at fault and name the responsible, malicious
party.

\textbf{Scope.}
Earlier works~\cite{yasaei2021gnn4tj,yu2021hw2vec,yasaei2022golden} are {limited to register transfer level (RTL)},
unable to handle gate-level
netlists (GLNs). Such 
methods are restricted to pre-silicon assessment; they cannot detect HTs in the
field. Note that only schemes which can work on GLNs allow for pre- and post-silicon detection.

\textbf{Associated Research Challenges.} Developing a GNN-based HT detection and localization scheme that can work on
GLNs imposes the following research challenges (RC).

\textit{RC1: GLN Complexity.}
Compared to RTL, GLN designs are more complex to analyze, as GLNs are flattened (i.e., hierarchical
information is lost) and also considerably larger, in the range of thousands or even millions of gates and wires.

\textit{RC2: Imbalanced Datasets.}
HTs are stealthy and small in size; 
HT gates
represent a very small percentage, e.g., 0.14--11.29\% or 1.94\% on average
for the TrustHub suite considered in
this work.
Thus, a highly imbalanced dataset arises (e.g., the ratio of regular to HT gates reaches up to 719$\times$ for
		the TrustHub suite),
 which is difficult to handle for any ML model.

\textbf{Our Contributions.} Here, we propose \textit{TrojanSAINT}, a GNN-based method for
HT detection and localization that works well on large-scale GLNs.
The concept is outlined in Fig.~\ref{fig:intro}.

As indicated, the graph representations of GLNs are complex and large, which makes them difficult to handle with traditional architectures such as graph convolutional networks (GCNs).

This motivates our decision to, without loss of generality (\wolog), use GraphSAINT~\cite{zeng2019graphsaint} for our
methodology. GraphSAINT is a well-established, sampling-based approach that extracts smaller sub-graphs for training from the
larger original graph. It has shown good performance for various tasks~\cite{zeng2019graphsaint,GNNUnlock, GNNUnlockp}, but it has not been
considered for HT detection until now. We summarize our contributions as follows:
\begin{enumerate}
\item A parser for GLN-to-graph conversion (Sec.~\ref{sec:parse}) which performs feature extraction tailored for HT detection. 
\item A GNN-based method for detection and localization of HT in GLNs (Sec.~\ref{sec:gnn}), addressing \textit{RC1}.
\item A procedure for tuning of the classification thresholds to obtain more accurate predictions,
	addressing \textit{RC2}.

\item We
demonstrate that our scheme is competitive to traditional ML baselines and prior art. 
We also verify the generalization ability of our scheme -- i.e., good prediction accuracy for unknown HTs on unseen
GLNs.
\item We open-source our scheme and related artifacts from our experimental study [\url{https://github.com/DfX-NYUAD/TrojanSAINT}]. 
\end{enumerate}

\section{TrojanSAINT Methodology}
\label{sec:method}

\begin{figure}[tb]
\centerline{\includegraphics[width=\columnwidth]{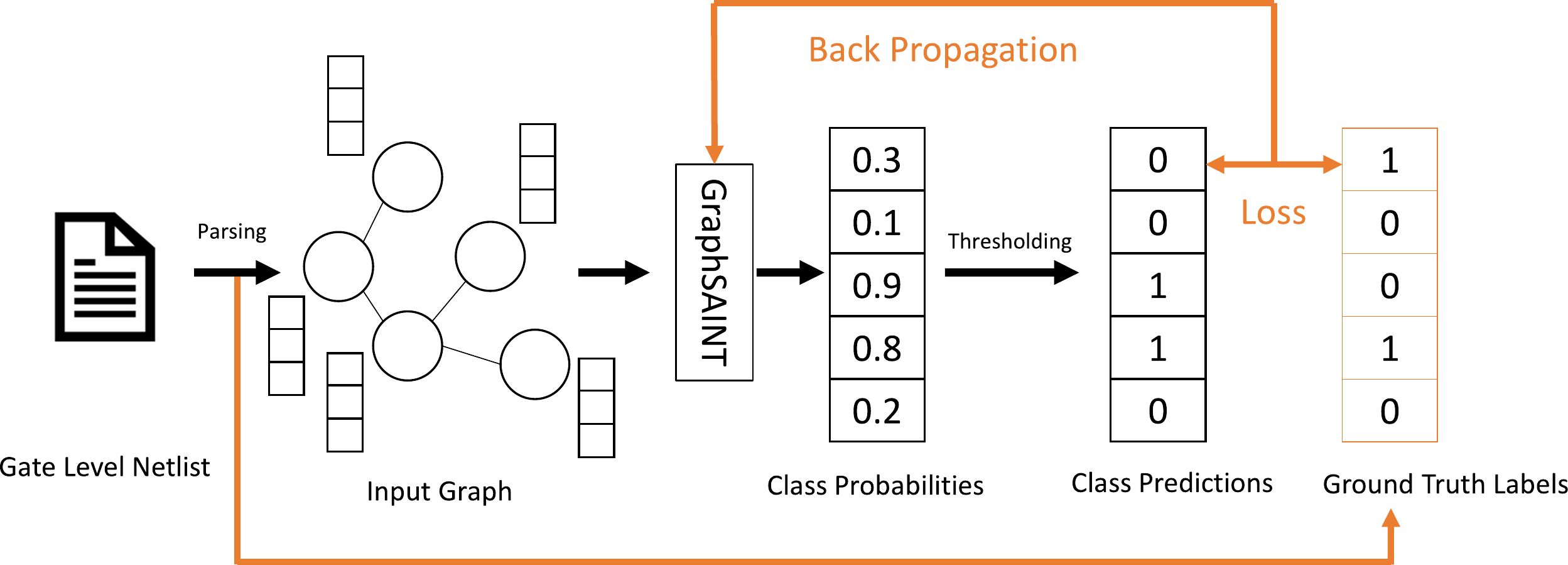}}
\caption{Overview of TrojanSAINT. Black arrows follow the inference process, orange arrows follow the
	additional steps needed for training and validation. In this example, the thresholding value is 0.4.}
\label{fig:method}
\end{figure}

An overview of our methodology is shown in Fig.~\ref{fig:method}. Next, we describe all relevant details.

\subsection{GLN Parsing and Feature Vectors}
\label{sec:parse}

\textbf{Parser.}
We develop a parser that converts GLNs (given in Verilog format) into unweighted and undirected graphs, where nodes
represent gates and edges represent wires. We are discarding directionality for improved representation
learning~\cite{GNNRE}. Given a set of GLNs, our parser generates one large single graph,
consisting of multiple disjoint graphs, where
nodes are labeled as `train,' `validation' or `test,' depending on the designation of the GLN they belong to.
The graph is
encoded as an adjacency matrix $\mA$ following a standard procedure.

\textbf{Feature Vectors.}
Our parser also generates a matrix $\mX$ of feature vectors for all nodes.
Vectors cover the following:
\begin{itemize}
 \item Gate type, represented via one-hot encoding. From experimentation, we are more interested in the functionality of the gate over the exact implementation.
 That is, we group functionally related gates together, e.g., all AND gates are grouped
 regardless of the number of inputs and the driver strengths that the different AND gates support.
 \item Input, output degrees of gates, i.e., the number of incoming and outgoing connections.
 \item Shortest distances to primary inputs/outputs.
 For gates not directly connected with
 a primary input/output, a breadth-first search is conducted to obtain shortest distances.

\end{itemize}
For training and validation, we also use a binary label vector which marks each
node as part of some HT or as regular/benign gate.
The related information is derived during parsing.

\subsection{GNN Implementation and Application}
\label{sec:gnn}

\textbf{Outline.} We utilize GraphSAINT~\cite{zeng2019graphsaint} for sampling. Further, we utilize the GNN architecture
of GNN-RE~\cite{alrahis2021gnn}, along with GraphSAGE~\cite{hamilton2017inductive}. We tune the classification thresholds for more accurate predictions.
We further utilize a practical validation.
For training and inference, we employ standard procedures.

\textbf{GNN Architecture.}
We consider an undirected graph 
$\G\paren{\V, \mA}$ for representing a GLN, where $\V$ is the set of vertices/nodes/gates,
and the adjacency matrix of the graph is $\mA$, where $A_{u,v}=1$ and $A_{v,u}=1$ if there exists an edge/wire from vertex/gate $u$ to
vertex/gate $v$. Each vertex $u$ in the initial graph $\G$ has a feature vector $\bm{x}_u$. This vector represents the node embedding at
layer zero of the GNN. The embedding of node $u$ is iteratively updated by the GNN, by aggregating the embedding of the node and its
neighbors $\N(u)$. The embedding of a node $u$ after $l$ GNN layers, $\bm{h}_u^{(l)}$, is given by:
\begin{align}
\footnotesize
 \bm{a}_u^{(l)} = \text{AGGREGATE}^{(l)} \left( \left\lbrace \bm{h}_v^{(l-1)} : v \in \mathcal{N}(u) \right\rbrace \right) 
\vspace*{-5mm}
\end{align}
 \vspace*{-5mm}
\begin{align}
\footnotesize
 \bm{h}_u^{(l)} = \text{COMBINE}^{(l)} \left( \bm{h}_u^{(l-1)}, \bm{a}_u^{(l)} \right)
 \vspace*{-5mm}
\end{align}

GNN architectures are defined by their implementation of \texttt{AGGREGATE}$(\cdot)$ and
\texttt{COMBINE}$(\cdot)$. For example, GraphSAGE~\cite{hamilton2017inductive}, which we also use here,
works as follows:

\begin{gather}
\bm{h}_{u}^{(l)}=\sigma ([\bm{W}_l\cdot AGG(\{\bm{h}_v^{(l-1)},\forall v\in \N(u)\}),\bm{B}_l \bm{h}_u^{(l-1)}])\\
AGG=\sum_{v\in \N(u)}\frac{\bm{h}_{v}^{(l-1)}}{|\N(u)|}
\end{gather}

where $\sigma(.)$ is an activation function such as \texttt{ReLU} and $\bm{W}_{l}$ and $\bm{B}_{l}$ are trainable weight matrices. In GraphSAGE, the embedding of node $\bm{h}_{u}^{(l)}$ is determined by first concatenating the node's features from
the previous layer $\bm{h}_{u}^{(l-1)}$ with the output of the AGG function. Then the $\bm{W}_{l}$ and $\bm{B}_{l}$
transformations learns the important components of the neighbors' features and the node $u$, respectively.
GraphSAGE is compatible with different $AGG$ functions. Here, we use the mean aggregator as described in Equation~(4).

\textbf{Thresholding.}
From experimentation, we observe that the classification
threshold plays a significant role for prediction performance. This is because of the considerably imbalanced datasets (Sec.~\ref{sec:intro}, RC2), where the GNN model as is
can predict the minority class, i.e., HT nodes, only with low confidence.

{The goal of thresholding is to determine a sufficiently small value so that HT nodes/gates
are classified as such the moment the GNN captures any hint of malicious structures. In other words, thresholding allows the GNN to focus more on the minority class, improving
the performance of the entire model.}

\Wolog, we tune the
threshold between 0--0.5 in steps of 1,000 and select the threshold
that yields the best score on validation. Here, best score refers to the average of true positive rate (TPR) and true
negative rate (TNR).

\textbf{Practical Validation.} 
We propose an approach where predictions are made on unknown HTs residing within circuits that are neither seen during
training nor have golden references. This
represents a real-world scenario, where security engineers do not know in advance which HT to expect, if any at all, and further need to test 
circuits without golden references.
Prior art did not necessarily consider such practical validation.

\textbf{Training.} First, we construct sub-graphs using a standard \textit{random-walk sampler} ({\rws}). TrojanSAINT's training procedure is shown in Algorithm~\ref{algo:gsaint_meta}.
Due to the \rws, the network can become biased towards frequently sampled nodes. To alleviate
this issue, we follow the normalization technique of~\cite{zeng2019graphsaint}.
We use stochastic gradient descent as optimizer. ${\G}_s$ is sampled
for each minibatch and a GNN is built on the sub-graph. The cross-entropy loss is calculated for each node in the
sub-graph and the GNN weights are then updated by backpropagation.

\begin{algorithm}[tb]
\caption{TrojanSAINT training algorithm}
\label{algo:gsaint_meta}
\begin{algorithmic}[1]
\footnotesize
\renewcommand{\algorithmicrequire}{\textbf{Input:}}
\renewcommand{\algorithmicensure}{\textbf{Output:}}
\Require Training graph $\G\paren{\V, \mA}$; Ground truth $\overline{\bm{Y}}$; Sampler \rws
\Ensure Trained GNN 
\State Compute normalization coefficients $\alpha$, $\lambda$ using {\rws}
\For{each mini-batch}
 \State ${\G}_s\paren{{\V}_s, {\mA}_s} \gets$ Sampled sub-graph of $\G$ using {\rws}
 \State Build GNN on ${\G}_s$
 \State $\set*{\bm{y}_u\given u\in \V_{s}}\gets$ Propagating $\alpha$-normalized $\set*{\bm{x}_u\given u\in \V_{s}}$
 \State Propagating $\lambda$-normalized loss $L\paren{\bm{y}_u,\overline{\bm{y}}_u}$ to update weights
\EndFor
\end{algorithmic}
\end{algorithm}

\textbf{Inference.}
See Algorithm~\ref{algo:gsaint_test}.
For all test nodes in the graph, node embeddings are calculated and passed to a
fully-connected layer with \texttt{softmax} activation, to compute class probabilities. We then apply our
thresholding technique, and finally convert class probabilities into labels.

\begin{algorithm}[tb]
\caption{TrojanSAINT inference algorithm}
\label{algo:gsaint_test}
\begin{algorithmic}[1]
\footnotesize
\renewcommand{\algorithmicrequire}{\textbf{Input:}}
\renewcommand{\algorithmicensure}{\textbf{Output:}}
\Require Flattened netlist $N$; Trained GNN; Threshold $th$
\Ensure Trojan classification of all nodes/gates
\State Initiate $\G\paren{\V, \mA}$ with $V \gets GLN\_to\_graph(N)$ 
\For{each $u \in \V$}
 \State $\bm{z}_{u} \gets GNN(u)$ \Comment Compute embedding
 \State $c_{u} \gets fc({z}_{u}, th)$ \Comment Classify node $u$ based on the threshold 
\EndFor
\end{algorithmic}
\end{algorithm}

\section{Experimental Study}
\label{sec:results}

\begin{table*}[tb]
\centering
\caption{TPR/TNR Results for Practical Validation. 
		Best Results, Considering Average of TPR and TNR, are Marked in Boldface.}

\smallerspacecaption
\label{tab:resultsnew}
\label{tab:resultsold}
\resizebox{.99\textwidth}{!}{%
\begin{tabular}{cccccccccccccccc}\toprule
\multirow{2}{*}{TrustHub} & \multicolumn{7}{c}{All With Thresholding} & & \multicolumn{7}{c}{Others Without Thresholding}\\ \cmidrule{2-8}\cmidrule{10-16}
{Benchmark} & \textbf{TrojanSAINT} & XGBoost & FCNN & GNN-RE & Logistic & Random & SVM & & \textbf{TrojanSAINT} & XGBoost & FCNN & GNN-RE & Logistic & Random & SVM \\
& & & & & Regression & Forest & 							& & & & & & Regression &
Forest & 					\\ \midrule 
rs232t1000 &\textbf{1.00/0.60} &1.00/0.57 &0.85/0.64 &0.77/0.51 &1.00/0.46 &0.69/0.75 &N/A & &\textbf{1.00/0.60} &0.23/0.91 &0.00/1.00 &0.00/1.00 &0.00/1.00 &0.15/0.92 &0.00/1.00 \\\cmidrule{2-8}\cmidrule{10-16}
rs232t1100 &0.92/0.68 &0.83/0.57 &\textbf{1.00/0.61} &0.92/0.52 &0.83/0.61 &0.33/0.90 &N/A & &\textbf{0.92/0.68} &0.08/0.91 &0.00/0.94 &0.00/0.94 &0.00/1.00 &0.08/0.92 &0.00/1.00 \\\cmidrule{2-8}\cmidrule{10-16}
rs232t1200 &0.41/0.80 &0.59/0.56 &\textbf{0.59/0.91} &0.82/0.27 &0.71/0.60 &0.35/0.75 &N/A & &\textbf{0.41/0.80} &0.06/0.90 &0.06/1.00 &0.06/0.93 &0.00/1.00 &0.00/0.91 &0.00/1.00 \\\cmidrule{2-8}\cmidrule{10-16}
rs232t1300 &\textbf{1.00/0.74} &1.00/0.57 &1.00/0.66 &1.00/0.71 &1.00/0.43 &0.56/0.86 &N/A & &\textbf{1.00/0.74} &0.22/0.91 &0.00/1.00 &0.00/0.98 &0.00/1.00 &0.00/0.92 &0.00/1.00 \\\cmidrule{2-8}\cmidrule{10-16}
rs232t1400 &0.92/0.50 &0.92/0.57 &\textbf{1.00/0.56} &1.00/0.27 &1.00/0.46 &0.54/0.70 &N/A & &\textbf{0.92/0.50} &0.08/0.91 &0.08/0.71 &0.62/0.94 &0.00/1.00 &0.00/0.92 &0.00/1.00 \\\cmidrule{2-8}\cmidrule{10-16}
rs232t1500 &\textbf{0.71/0.82} &0.93/0.57 &0.93/0.57 &0.79/0.61 &1.00/0.46 &0.57/0.76 &N/A & &\textbf{0.71/0.82} &0.21/0.91 &0.07/0.77 &0.36/0.94 &0.00/1.00 &0.14/0.91 &0.00/1.00 \\\cmidrule{2-8}\cmidrule{10-16}
rs232t1600 &0.73/0.57 &0.73/0.57 &\textbf{0.91/0.49} &0.55/0.66 &0.73/0.59 &0.27/0.90 &N/A & &\textbf{0.73/0.57} &0.18/0.91 &0.00/1.00 &0.00/0.83 &0.00/1.00 &0.00/0.91 &0.00/1.00 \\\cmidrule{2-8}\cmidrule{10-16}
s15850t100 &0.35/0.97 &0.77/0.94 &0.92/0.76 &\textbf{0.88/0.97} &0.96/0.73 &0.85/0.94 &N/A & &\textbf{0.35/0.97} &0.12/1.00 &0.00/1.00 &0.12/1.00 &0.00/1.00 &0.04/1.00 &0.04/1.00 \\\cmidrule{2-8}\cmidrule{10-16}
s35932t100 &\textbf{1.00/1.00} &0.87/0.98 &0.87/0.64 &0.93/1.00 &0.93/0.44 &1.00/0.98 &N/A & &\textbf{1.00/1.00} &0.20/1.00 &0.00/1.00 &0.00/0.97 &0.00/1.00 &0.13/1.00 &0.07/1.00 \\\cmidrule{2-8}\cmidrule{10-16}
s35932t200 &\textbf{1.00/1.00} &1.00/0.98 &0.92/0.80 &\textbf{1.00/1.00} &0.92/0.44 &1.00/0.99 &N/A & &\textbf{1.00/1.00} &0.00/1.00 &0.00/1.00 &0.00/1.00 &0.00/1.00 &0.00/1.00 &0.00/1.00 \\\cmidrule{2-8}\cmidrule{10-16}
s35932t300 &0.97/1.00 &0.94/0.98 &1.00/0.81 &\textbf{1.00/1.00} &0.40/0.81 &0.97/0.97 &N/A & &\textbf{0.97/1.00} &0.63/1.00 &0.00/1.00 &0.09/1.00 &0.00/1.00 &0.57/1.00 &0.00/1.00 \\\cmidrule{2-8}\cmidrule{10-16}
s38417t100 &0.92/0.92 &1.00/0.82 &0.75/0.77 &\textbf{1.00/0.92} &1.00/0.35 &0.75/0.90 &N/A & &\textbf{0.92/0.92} &0.33/0.95 &0.00/1.00 &0.00/1.00 &0.00/1.00 &0.42/0.94 &0.00/1.00 \\\cmidrule{2-8}\cmidrule{10-16}
s38417t200 &0.40/0.99 &0.53/0.86 &0.73/0.73 &0.47/0.93 &1.00/0.35 &\textbf{0.73/0.90} &N/A & &\textbf{0.40/0.99} &0.27/0.95 &0.73/0.90 &0.00/1.00 &0.00/1.00 &0.27/0.94 &0.00/1.00 \\\cmidrule{2-8}\cmidrule{10-16}
s38417t300 &\textbf{0.98/0.96} &0.98/0.82 &0.18/0.89 &0.98/0.91 &0.16/0.87 &0.95/0.84 &N/A & &\textbf{0.98/0.96} &0.14/0.95 &0.07/1.00 &0.00/0.98 &0.02/1.00 &0.23/0.95 &0.07/1.00 \\\cmidrule{2-8}\cmidrule{10-16}
s38584t100 &\textbf{1.00/0.95} &1.00/0.87 &1.00/0.87 &1.00/0.92 &1.00/0.52 &1.00/0.93 &N/A & &\textbf{1.00/0.95} &0.22/1.00 &0.00/1.00 &0.00/1.00 &0.00/1.00 &0.22/1.00 &0.00/1.00 \\\cmidrule{2-8}\cmidrule{10-16}
s38584t200 &\textbf{0.90/0.98} &0.49/0.87 &0.89/0.87 &0.39/0.95 &0.98/0.52 &0.84/0.94 &N/A & &\textbf{0.90/0.98} &0.02/1.00 &0.00/1.00 &0.00/1.00 &0.00/1.00 &0.02/1.00 &0.02/1.00 \\\cmidrule{2-8}\cmidrule{10-16}
s38584t300 &0.13/0.98 &0.08/0.88 &0.47/0.94 &0.23/0.93 &\textbf{0.94/0.52} &0.45/0.94 &N/A & &\textbf{0.13/0.98} &0.01/1.00 &0.00/1.00 &0.00/1.00 &0.00/1.00 &0.01/1.00 &0.00/1.00 \\\cmidrule{2-8}\cmidrule{10-16}
\textit{Average} &\textbf{0.78/0.85} &0.80/0.76 &0.82/0.74 &0.81/0.77 &0.86/0.54 &0.70/0.88 &N/A & &\textbf{0.78/0.85} &0.18/0.95 &0.06/0.96 &0.07/0.97 &0.00/1.00 &0.13/0.95 &0.01/1.00	\\ 
\bottomrule
\end{tabular}
}
\smallerspacecaption
\smallerspacecaption
\end{table*}

\subsection{Setup}
\label{sec:setup}

\textbf{Software.}
We use Python for coding and bash scripts for job/data management.
TrojanSAINT extends on
GNN-RE, which is obtained from~\cite{GNNRE} and is
implemented in PyTorch. Our TrojanSAINT platform is available online [\url{https://github.com/DfX-NYUAD/TrojanSAINT}]. Baseline models are implemented using Scikit-Learn, except the fully-connected neural network (FCNN) in
PyTorch.

\textbf{Computation.}
Experiments for GNN-RE, TrojanSAINT and FCNN are conducted on a high-performance cluster with 4x
Nvidia V100 GPUs and 360GB RAM;
experiments for others are conducted on a workstation with Intel i7 CPU and 16GB RAM.
Training of GNN-RE and TrojanSAINT takes $\approx$15--30 minutes per model, FCNN $\approx$10 minutes per model, and all others
$\approx$3 minutes in total. All inference takes few seconds.

\textbf{Benchmarks and Model Building.}
We use 17 exemplary GLN benchmarks from the TrustHub suite~\cite{TH}. 
For each benchmark, a respective model is trained from scratch. For our practical validation, each model does not get to
see the design to be tested at all during training.\footnote{%
For example, if rs232t1000 is to be tested, {none of the other rs232
designs are used for training, only for validation.}

For s15850t100, the only s15850 design in the suite, we randomly select three other designs for
validation.}

We note that random seeds used in TrojanSAINT's components affect performance significantly. Thus, we conduct \wolog\ 6 runs with different seeds and report only results for each model that performs best on its validation set.

\textbf{Prior Art, Comparative Study.}
From Table~\ref{tab:comparison}, recall that none of the prior art in GNN-based HT detection works on GLNs.
Thus, a direct comparison is not practical.
However, we consider the following works for comparison.
\begin{itemize}
\item GNN-RE~\cite{alrahis2021gnn}: Proposed for reverse engineering of GLNs, it could also be utilized for HT
detection and localization.
This is because GNN-RE seeks to classify gates/nodes from flattened GLNs into the circuit modules they belong to;
TrojanSAINT's task of classifying gates/nodes into begin or HT-infested ones is analogous.

\item Related Works~\cite{r-ht,kento17,tatsuki21}: ML-based, not GNN-based, HT detection schemes that are working 
on GLNs. Unlike ours, these works employ elaborate feature engineering. Also, these works
do not offer native HT localization.

\end{itemize}

We also implement and run the following well-known baseline classifiers for a further comparative study.
\begin{itemize}
 \item XGBoost: A decision tree (DT)-based model that uses an ensemble of sequentially added DTs. DTs are
 added aiming to minimize errors of their predecessor.
 \item Random Forest: A DT-based model that uses an ensemble of DTs trained on subsets of the
 training data.
 \item Logistic Regression: A classification algorithm that utilizes the \texttt{sigmoid} function on
 independent variables.
 \item Support Vector Machine (SVM): A classification model that generates a hyperplane to separate different classes.
 \item %
	 FCNN: We implement a three-layer network; each layer use the
 \texttt{SELU} activation function~\cite{klambauer2017self} and batch normalization. The final layer uses \texttt{sigmoid} activation
 to calculate classification probabilities.
\end{itemize}
All these classifiers work on tabular, non-graph data; thus, we provide them with the feature vectors
as inputs.
All classifiers, except SVM, output probabilities; thus, we can study them considering our proposed thresholding as
well.

\subsection{Results}

\textbf{Practical Validation and Impact of Thresholding.}
In Table~\ref{tab:resultsnew}, we report
TPR/TNR results for practical validation across two scenarios: with thresholding versus without.\footnote{
Since thresholding is part of our proposed scheme, we do not consider TrojanSAINT without.
We implement the same thresholding strategy (Sec.~\ref{sec:gnn}) for all models. 
SVM directly separates data into classes without computing probabilities, making thresholding not applicable
	(N/A).}

First, the results show that TrojanSAINT outperforms
other methods
for this realistic but challenging scenario
of HT detection considering unknown Trojans within unseen circuits.
The GNN framework underlying of TrojanSAINT is superior to other models. Recall that others take the
same feature vectors as inputs; such direct comparison is fair.
Second, thresholding is crucial for high prediction performance for this task.

\textbf{Relaxed Validation.} We also study a ``best case'' validation,
using a leave-one-out split where validation and test sets are the same. Such setting is often considered in the literature, as it shows the best performance for any model and benchmark.
As indicated, however, it is not as realistic for HT detection.

With thresholding applied, we observe the following average TPR/TNR values here:\footnote{%
	Due to limited space, we refrain from reporting a table for this scenario.}
	{0.98/0.96} for TrojanSAINT, 0.93/0.93 for XGBoost, 0.91/0.89 for
FCNN, 0.98/0.96 for GNN-RE, 0.89/0.81 for logistic regression, and 0.91/0.994 for random forest, respectively.
Without thresholding applied, we observe the following average TPR/TNR values: 0.41/0.99 for XGBoost, 0.09/1.00 for
FCNN, 0.07/0.97 for GNN-RE, 0.09/1.00 for logistic regression, 0.40/0.99 for random forest, and 0.11/1.00 for SVM, respectively.
TrojanSAINT is superior to almost all methods across these two cases; only GNN-RE, and only with thresholding applied, becomes a close
contender.

\textbf{Related Works.}
In Table~\ref{tab:related}, we compare to more loosely related works (Sec.~\ref{sec:setup}). Results are quoted and rounded.
Numbers of nodes/gates are reported as obtained from our parser.\footnote{%
Number of nodes/gates may vary across ours and related works, depending on parsing approach, technology library etc., but 
overall ranges remain similar.} {The related works employ leave-one-out or ``best case'' validation schemes; thus, we also report
	TrojanSAINT results for such ``best case'' validation here.}

\begin{table}[tb]
\centering
\caption{Benchmark Properties; TPR/TNR Results for Related Works}
\smallerspacecaption
\label{tab:related}
\resizebox{\columnwidth}{!}{%
\begin{tabular}{ccccccccc}\toprule
{TrustHub} &Benign &HT &Ratio of Nodes,& & {R-HTD~\cite{r-ht}} & \multirow{2}{*}{\cite{kento17}} &
\multirow{2}{*}{\cite{tatsuki21}} & \multirow{2}{*}{\textbf{TrojanSAINT}} \\ 
{Benchmark} & Nodes & Nodes & HT to Benign & & (Orig. Samples) & & \\ \midrule 
rs232t1000 &202 &13 &0.064 & &1.00/0.94 &1.00/0.99 &\textbf{1.00/1.00} &1.00/0.94 \\ \cmidrule{2-4}\cmidrule{6-9}
rs232t1100 &204 &12 &0.059 & &1.00/0.93 &0.50/0.98 &\textbf{1.00/1.00} &1.00/0.93 \\ \cmidrule{2-4}\cmidrule{6-9}
rs232t1200 &199 &17 &0.085 & &0.97/0.96 &0.88/1.00 &\textbf{1.00/1.00} &0.82/0.96 \\ \cmidrule{2-4}\cmidrule{6-9}
rs232t1300 &204 &9 &0.044 & &1.00/0.95 &\textbf{1.00/1.00} &0.86/1.00 &1.00/0.98 \\ \cmidrule{2-4}\cmidrule{6-9}
rs232t1400 &202 &13 &0.064 & &1.00/0.98 &0.98/1.00 &\textbf{1.00/1.00} &1.00/0.96 \\ \cmidrule{2-4}\cmidrule{6-9}
rs232t1500 &202 &14 &0.069 & &1.00/0.94 &0.95/1.00 &\textbf{1.00/1.00} &1.00/0.94 \\ \cmidrule{2-4}\cmidrule{6-9}
rs232t1600 &203 &11 &0.054 & &\textbf{0.97/0.92} &0.93/0.99 &0.78/0.99 &1.00/0.88 \\ \cmidrule{2-4}\cmidrule{6-9}
s15850t100 &2,156 &26 &0.012 & &0.74/0.93 &0.78/1.00 &0.08/1.00 &\textbf{0.88/0.97} \\ \cmidrule{2-4}\cmidrule{6-9}
s35932t100 &5,426 &15 &0.003 & &0.80/0.69 &0.73/1.00 &0.08/1.00 &\textbf{1.00/0.97} \\ \cmidrule{2-4}\cmidrule{6-9}
s35932t200 &5,426 &12 &0.002 & &0.08/1.00 &0.08/1.00 &0.08/1.00 &\textbf{1.00/1.00} \\ \cmidrule{2-4}\cmidrule{6-9}
s35932t300 &5,427 &35 &0.006 & &0.84/1.00 &0.81/1.00 &0.92/1.00 &\textbf{1.00/1.00} \\ \cmidrule{2-4}\cmidrule{6-9}
s38417t100 &5,329 &12 &0.002 & &0.67/1.00 &0.33/1.00 &0.09/1.00 &\textbf{1.00/0.97} \\ \cmidrule{2-4}\cmidrule{6-9}
s38417t200 &5,329 &15 &0.003 & &0.73/0.99 &0.47/1.00 &0.09/1.00 &\textbf{1.00/0.97} \\ \cmidrule{2-4}\cmidrule{6-9}
s38417t300 &5,329 &44 &0.008 & &0.89/1.00 &0.75/1.00 &\textbf{1.00/1.00} &1.00/0.96 \\ \cmidrule{2-4}\cmidrule{6-9}
s38584t100 &6,473 &9 &0.001 & &N/A &N/A &0.17/1.00 &\textbf{1.00/0.99} \\ \cmidrule{2-4}\cmidrule{6-9}
s38584t200 &6,473 &83 &0.013 & &N/A &N/A &0.18/1.00 &\textbf{1.00/0.98} \\ \cmidrule{2-4}\cmidrule{6-9}
s38584t300 &6,473 &731 &0.113 & &N/A &N/A &0.03/1.00 &\textbf{0.99/0.95} \\ \cmidrule{2-4}\cmidrule{6-9}
\multirow{2}{*}{\textit{Average}} & \multirow{2}{*}{3,250} & \multirow{2}{*}{63} & 0.035$^*$ & &
\multirow{2}{*}{0.84/0.95} & \multirow{2}{*}{0.72/1.00} & \multirow{2}{*}{0.55/1.00} &
\multirow{2}{*}{\textbf{{0.98/0.96}}} \\
& & & 0.019$^*$ & & & & & \\
\bottomrule
\end{tabular}
}\\[0.5mm] 
\scriptsize
\begin{justify}
$^*$The first value is averaged across the column; the second value, more representative of the overall
imbalance, is based on re-calculating the ratio
using the average node counts.

\end{justify}
\smallerspacecaption
\smallerspacecaption
\smallerspacecaption
\end{table}

TrojanSAINT outperforms these related works for all larger benchmarks, where the ratio of HT
gates/nodes to regular ones is more challenging---this demonstrates
superior scalability for ours.
For the smaller benchmarks, which are not representative of real IC designs, related works achieve better
results presumably due to feature engineering. In fact, up to 76 features are considered in~\cite{kento17,tatsuki21}
which reflects on considerable efforts, whereas for ours, some simple feature vectors suffice.

\section{Conclusion}
\label{sec:conc}

We have developed TrojanSAINT, a GNN-based method for detection and localization of HTs. We overcome the HT-inherent issue of class
imbalance through threshold tuning. Through practical validation, ours is capable of
generalizing to circuits and HTs it has not seen for training. Our method outperforms prior art and a number of strong ML baselines.
The use of a GNN framework renders TrojanSAINT simple yet competitive.
For future work, we will study the role of different feature vectors in more details.

\bibliographystyle{IEEEtran}
\bibliography{main}

\end{document}

%% file: settings.tex
\IEEEoverridecommandlockouts
\usepackage{cite}
\usepackage{amsmath,amssymb,amsfonts}
\usepackage{graphicx}
\usepackage{textcomp}
\usepackage{xcolor}

\usepackage{comment}

\newcommand{\wolog}{w/o.l.o.g.}
\newcommand{\Wolog}{W/o.l.o.g.}

\usepackage{cite}

\usepackage{blindtext,graphicx}
\usepackage{tikz}
\usepackage{relsize}
\usepackage{comment}
\usetikzlibrary{snakes}
\usepackage{amsmath,amssymb,amsfonts}
\usepackage{bm}
\usepackage{textcomp}
\usepackage{comment}

\usepackage{subfigure}
\usepackage{ragged2e}

\usepackage{balance}
\usepackage{mathtools}

\usepackage[caption=false,font=footnotesize]{subfig}
\usepackage{ifthen}
\usepackage{tikz}
\usepackage{siunitx}
\usepackage{xstring}
\usepackage{tikz}

\usepackage{pgfplots}
\definecolor{cadmiumgreen}{rgb}{0.0, 0.42, 0.24}

\usepackage{fancyhdr}
\usepackage{graphicx}
\usepackage{url}
\usepackage{multirow}
\usepackage{hhline}
\usepackage[us]{datetime}
\usepackage{paralist}
\usepackage{color}
\usepackage{soul}
\definecolor{cadmiumgreen}{rgb}{0.0, 0.42, 0.24}
\makeatletter
\def\blfootnote{\xdef\@thefnmark{}\@footnotetext}
\makeatother

\usepackage{pifont}
\usepackage{fancyhdr}

\input{math_commands}

\newcommand{\rws}{{\fontfamily{lmtt}\selectfont RWS}}
\usepackage{algorithm}
\usepackage{algorithmicx}
\usepackage{algpseudocode}
\usepackage{mathtools}
\usepackage[inline]{enumitem} 
\usepackage{pgfplots}

\usepackage{nicefrac}  
\usepackage{microtype}      
\usepackage{multirow}
\usepackage{amsthm,bbm}
\usepackage{ctable}
\usepackage{xstring}
\usepackage{tikz}

\usepackage{textcomp}

\usepackage{verbatim}
\usepackage{stmaryrd}
\usepackage{url}
\usepackage{bm}
\usepackage{graphicx}
\usepackage{amsmath,amsfonts,amssymb,amsthm}
\usepackage{mathtools}
\DeclarePairedDelimiterX\set[1]\lbrace\rbrace{\def\given{\;\delimsize\vert\;}#1}

%% file: math_commands.tex
\usepackage{amsmath,amsfonts,bm}

\newcommand{\G}[1][]{
    \IfEq{#1}{}
    {\mathcal{G}}
    {\mathcal{G}}_{#1}}
\newcommand{\V}[1][]{
    \IfEq{#1}{}
    {\mathcal{V}}
    {\mathcal{V}_{#1}}}
\newcommand{\E}[1][]{
    \IfEq{#1}{}
    {\mathcal{E}}
    {\mathcal{E}_{#1}}}
\newcommand{\N}[1][]{
    \IfEq{#1}{}
    {\mathcal{N}}
    {\mathcal{N}}_{#1}}
\newcommand{\X}[1][]{
    \IfEq{#1}{}
    {\bm{X}}
    {\bm{X}^{\paren{#1}}}}
\newcommand{\Asym}[1][]{
    \IfEq{#1}{}
    {\widetilde{\bm{A}}}
    {\widetilde{\bm{A}}_{#1}}}
\newcommand{\Arw}[1][]{
    \IfEq{#1}{}
    {\widehat{\bm{A}}}
    {\widehat{\bm{A}}_{#1}}}
\newcommand{\A}[1][]{
    \IfEq{#1}{}
    {\bm{A}}
    {\bm{A}}_{#1}}
\newcommand{\Hx}[1][]{
    \IfEq{#1}{}
    {\bm{H}}
    {\bm{H}}^{(#1)}}

\newcommand{\W}[1][]{
    \IfEq{#1}{}
    {\bm{W}}
    {\bm{W}^{\paren{#1}}}}

\newcommand{\paren}[1]{\left( #1 \right)}

\def\mA{{\bm{A}}}

\def\mX{{\bm{X}}}

\DeclareMathAlphabet{\mathsfit}{\encodingdefault}{\sfdefault}{m}{sl}
\SetMathAlphabet{\mathsfit}{bold}{\encodingdefault}{\sfdefault}{bx}{n}

%% file: authors.tex
\author{
\IEEEauthorblockN{Hazem Lashen, Lilas Alrahis, Johann Knechtel, and Ozgur Sinanoglu}
\IEEEauthorblockA{New York University Abu Dhabi\\
\{hl3372, lma387, jk176, os22\}@nyu.edu}}

%% file: abstract.tex
We propose \textit{TrojanSAINT}, a graph neural network (GNN)-based hardware Trojan (HT) detection scheme working at the gate level. Unlike prior GNN-based art, TrojanSAINT enables both pre-/post-silicon HT detection. TrojanSAINT leverages a
sampling-based GNN framework to detect and also localize HTs. For practical validation, TrojanSAINT achieves on average (oa) 78\% true positive rate (TPR) and 85\% true negative rate
(TNR), respectively, on various TrustHub HT benchmarks.
For best-case validation, TrojanSAINT even achieves {98\% TPR and 96\% TNR} oa. TrojanSAINT outperforms related prior works and baseline classifiers. We release our source codes and result artifacts.